\documentclass{webofc}
\usepackage[varg]{txfonts}   
\usepackage{hyperref}
\usepackage{epstopdf}
\usepackage{mhchem}
\usepackage{upgreek}
\usepackage{comment}
\usepackage{xspace}
\usepackage{placeins}
\usepackage{lineno} 
\usepackage{placeins}  
\usepackage{float}

\hypersetup{colorlinks=true,citecolor=blue,urlcolor=blue,linkcolor=blue}

\newcommand{\pp}           {pp\xspace}

\newcommand{\PbPb}         {\mbox{Pb--Pb}\xspace}

\newcommand{\s}            {\ensuremath{\sqrt{s}}\xspace}

\newcommand{\pt}           {\ensuremath{p_{\rm T}}\xspace}

\newcommand{\RAA}         {\ensuremath{R_{\rm AA}}\xspace}

\newcommand{\nineH}        {$\sqrt{s}~=~0.9$~Te\kern-.1emV\xspace}
\newcommand{\seven}        {$\sqrt{s}~=~7$~Te\kern-.1emV\xspace}
\newcommand{\thirteen}        {$\sqrt{s}~=~13$~Te\kern-.1emV\xspace}
\newcommand{\thirteensix}{\ensuremath{\sqrt{s} = 13.6~\text{Te\kern-.1emV}}\xspace}
\newcommand{\twoH}         {$\sqrt{s}~=~0.2$~Te\kern-.1emV\xspace}
\newcommand{\twosevensix}  {$\sqrt{s}~=~2.76$~Te\kern-.1emV\xspace}
\newcommand{\five}         {$\sqrt{s}~=~5.02$~Te\kern-.1emV\xspace}
\newcommand{\twosevensixnn}{$\sqrt{s_{\mathrm{NN}}}~=~2.76$~Te\kern-.1emV\xspace}
\newcommand{\fivenn}       {$\sqrt{s_{\mathrm{NN}}}~=~5.02$~Te\kern-.1emV\xspace}
\newcommand{\fivethirtysix}       {$\sqrt{s_{\mathrm{NN}}}~=~5.36$~Te\kern-.1emV\xspace}

\newcommand{\GeVc}         {Ge\kern-.1emV/$c$\xspace}
\newcommand{\MeVc}         {Me\kern-.1emV/$c$\xspace}
\newcommand{\TeV}          {Te\kern-.1emV\xspace}
\newcommand{\GeV}          {Ge\kern-.1emV\xspace}
\newcommand{\MeV}          {Me\kern-.1emV\xspace}
\newcommand{\GeVmass}      {Ge\kern-.2emV/$c^2$\xspace}
\newcommand{\MeVmass}      {Me\kern-.2emV/$c^2$\xspace}

\newcommand{\ee}           {\ensuremath{\rm{e}^{+}\rm{e}^{-}}}

\newcommand{\dzero}        {\ensuremath{{\rm D}^{0}}\xspace}

\newcommand{\jpsi}         {\ensuremath{{\rm J}/\psi}\xspace}
\newcommand{\ds}{\ensuremath{\rm D^{+}_{s}}\xspace}
\newcommand{\dplus}{\ensuremath{\rm D^{+}}\xspace}
\newcommand{\sigfive}{\ensuremath{\Sigma^{0}_{c}(2520)}\xspace}
\newcommand{\sigfour}{\ensuremath{\Sigma^{0}_{c}(2455)}\xspace}

\newcommand{\dstar}{\ensuremath{\rm D^{*}}\xspace}
\newcommand{\Lc}{\ensuremath{\rm \Lambda_{c}}\xspace}

\newcommand{\psitwos}         {\ensuremath{\psi(\rm 2S)}\xspace}

\begin{document}
\title{ALICE highlights}

\author{\firstname{Xiaozhi} \lastname{Bai on behalf of the ALICE Collaboration. \\
		Email: \email{baixiaozhi@ustc.edu.cn}}} 
\institute{State Key Laboratory of Particle Detection and Electronics, \\
	University of Science and Technology of China, Hefei 230026, China}

\abstract{
The recent ALICE results on hard probes, focusing on open heavy flavor, quarkonia, and jet measurements in pp and Pb--Pb collisions are presented. During the LHC Run 3, the continuous readout of the upgraded Time Projection Chamber (TPC) with GEM technology, along with improved tracking from ITS2 and the newly installed Muon Forward Tracker (MFT), have enabled high-precision studies of quarkonia and heavy flavour production, and jet modification in the quark-gluon plasma (QGP). These new measurements provide crucial constraints on parton energy loss mechanisms and heavy-quark dynamics in hot QCD matter. Future upgrades planned for Run 4, as well as the long-term ALICE 3 with an all-silicon tracking system, will further enhance the ability to study the properties of QGP with unprecedented accuracy.
}
\maketitle

\section{Introduction}
\label{intro}

The ALICE experiment at the Large Hadron Collider (LHC) is designed to study the properties of strongly interacting matter at extreme temperatures and energy densities, where the formation of the quark-gluon plasma (QGP) is expected to occur~\cite{QGP_Review}. Understanding the QGP is essential for investigating the phase transition from ordinary to deconfined matter expected to happen in heavy-ion collisions. 

A fundamental approach for studying the QGP involves measurements of hard probes, including heavy-flavor hadrons, quarkonia, and jets. These probes are produced in the initial stages of the collision and traverse the QGP, undergoing modifications that encode information about the medium's properties. Heavy-flavor quarks, due to their large masses, serve as effective probes of the QGP transport properties and energy loss mechanisms~\cite{ALICE_Dmeson_2022}. Quarkonia provide insights into color screening effects and recombination processes in the medium~\cite{ALICE:2023gco}. Meanwhile, jets offer a direct way to study parton energy loss through interactions with the dense QCD medium~\cite{ALICE_Jet_Quenching_2023}.

In proton-proton (pp) collisions, the study of hard probe production serves as crucial baselines for testing perturbative QCD calculations, constraining parton distribution functions, and understanding heavy-quark fragmentation and hadronization~\cite{Frixione:2005yf,Bai:2024pxk,Braun-Munzinger:2015hba}.

The ALICE experiment has undergone major upgrades for Run 3, enhancing its capabilities to perform precision measurements of hard probes. A key advancement is the continuous readout of the Time Projection Chamber (TPC) based on Gas Electron Multiplier (GEM) technology~\cite{Hauer:2022kqo}, which enables the collection of large statistics by adopting a new data-taking paradigm. This upgrade significantly increases the recorded event statistics. Complementary to this, the improved Inner Tracking System (ITS2)~\cite{Liu:2024hlx} provides higher spatial resolution for vertex reconstruction. The newly installed Muon Forward Tracker (MFT) further increases the possibilities for studying heavy-flavor and quarkonium states at forward rapidity~\cite{Krupova:2024smm}.

Looking ahead, the Run 4 upgrades will include the replacement of the innermost layers of ITS2 with the ultra-thin ITS3, improving spatial resolution and reducing the material budget, which is particularly beneficial for tracking heavy-flavor hadrons and reconstructing displaced vertices. Additionally, the Forward Calorimeter (FoCal)~\cite{Inaba:2025boq} will enable precise measurements of direct photons and jets at forward rapidities, providing new constraints on small-$x$ gluon dynamics and initial-state effects. Beyond Run 4, the ALICE 3 upgrade aims to further improve vertex resolution, enhance particle identification, and enable high-precision jet reconstruction, more differential studies of heavy-quark transport, quarkonium suppression, and parton energy loss. These advancements will deepen our understanding of the QGP and the strong interaction under extreme conditions.

These proceedings present recent ALICE results on hard probes in pp and Pb--Pb collisions, emphasizing the impact of Run 3 upgrades. The findings provide new insights into the microscopic properties of the QGP and discuss upgrades for future Run 4 and beyond.

\FloatBarrier

\section{Hard probes in the small system}  

The study of hard probes in small collision systems, such as pp and p--Pb collisions~\cite{Cacciari:1998it,Vogt:2010}, provides essential insights into initial-state effects and the role of cold nuclear matter in high-energy interactions. While small systems are typically used as a reference for studying effects related to QGP production, pp collisions also serve as crucial tests of quantum chromodynamics (QCD). Hard probes, including heavy-flavor hadrons, quarkonia, and jets, provide direct access to both perturbative QCD (pQCD) processes and non-perturbative phenomena such as hadronization. Precision measurements in pp collisions constrain theoretical models and improve our understanding of charm and beauty quark production, fragmentation, and hadronization~\cite{ALICE:2019rmo}.

\begin{figure}[H]
	\centering
	\includegraphics[width=0.42\textwidth]{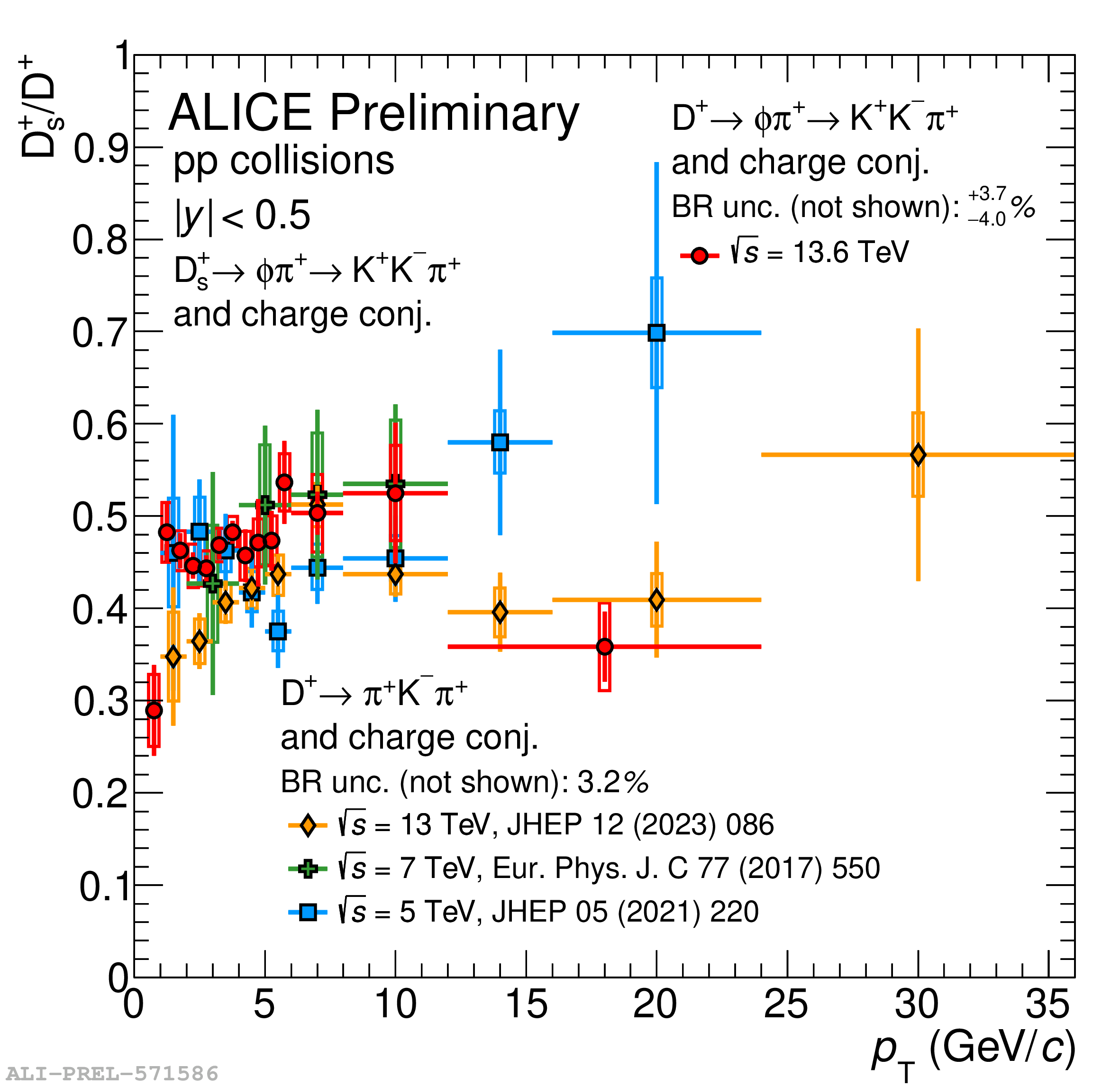}
	\includegraphics[width=0.52\textwidth]{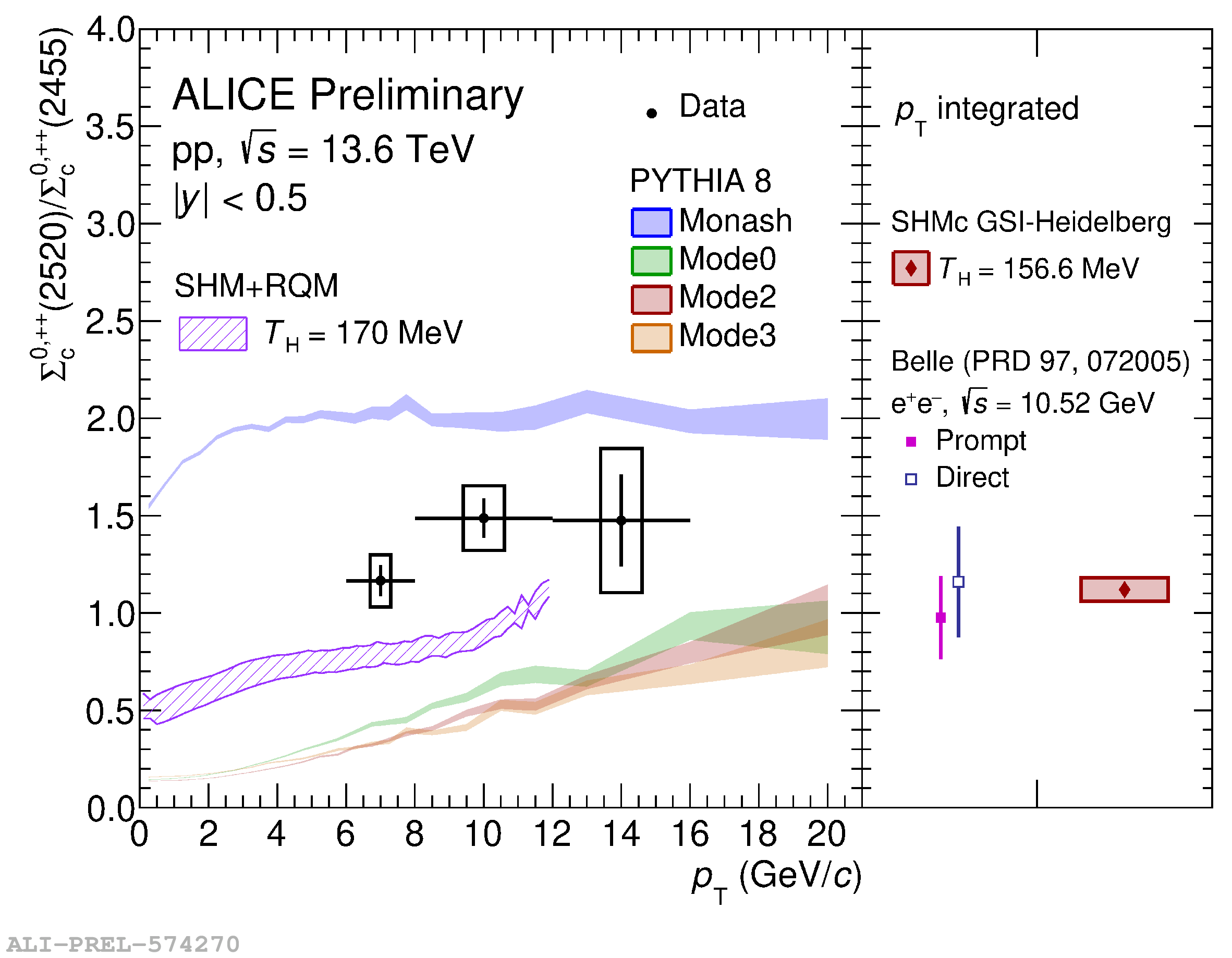}
	\caption{(Left) The ratio of \ds to \dplus as a function of \pt in \pp collisions at different energies. 
		(Right) The ratio of \sigfive to \sigfour as a function of \pt in \pp collisions at \thirteensix. 
		The results are compared with different model calculations, as well as similar measurements from the Belle experiment in \ee collisions at \s = 10.52 \GeV.}
	\label{openhfpp}
\end{figure}

The \pt-dependent yield ratio of $D_s^+$ to $D^+$ mesons in \pp collisions is shown in the left panel of Figure~\ref{openhfpp}. This new measurement extends the \pt coverage down to \pt~$= 0$. The results are compared at different center-of-mass energies, and no significant dependence on energy or \pt is observed. The yield ratio of excited charm baryons $\Sigma_c(2520)$ to $\Sigma_c(2455)$ as a function of \pt in \pp collisions at $\sqrt{s} = 13.6$~TeV is shown in the right panel of Figure~\ref{openhfpp}. The experimental data are compared with PYTHIA predictions using different tunings, as well as measurements from the Belle experiment in electron-positron ($e^+e^-$) collisions at $\sqrt{s} = 10.52$~GeV. The results cannot be described by PYTHIA predictions but are consistent with the \pt-integrated ratio measured in $e^+e^-$ collisions.

\begin{figure}[H]
 	\centering
 	\includegraphics[width=6.3cm]{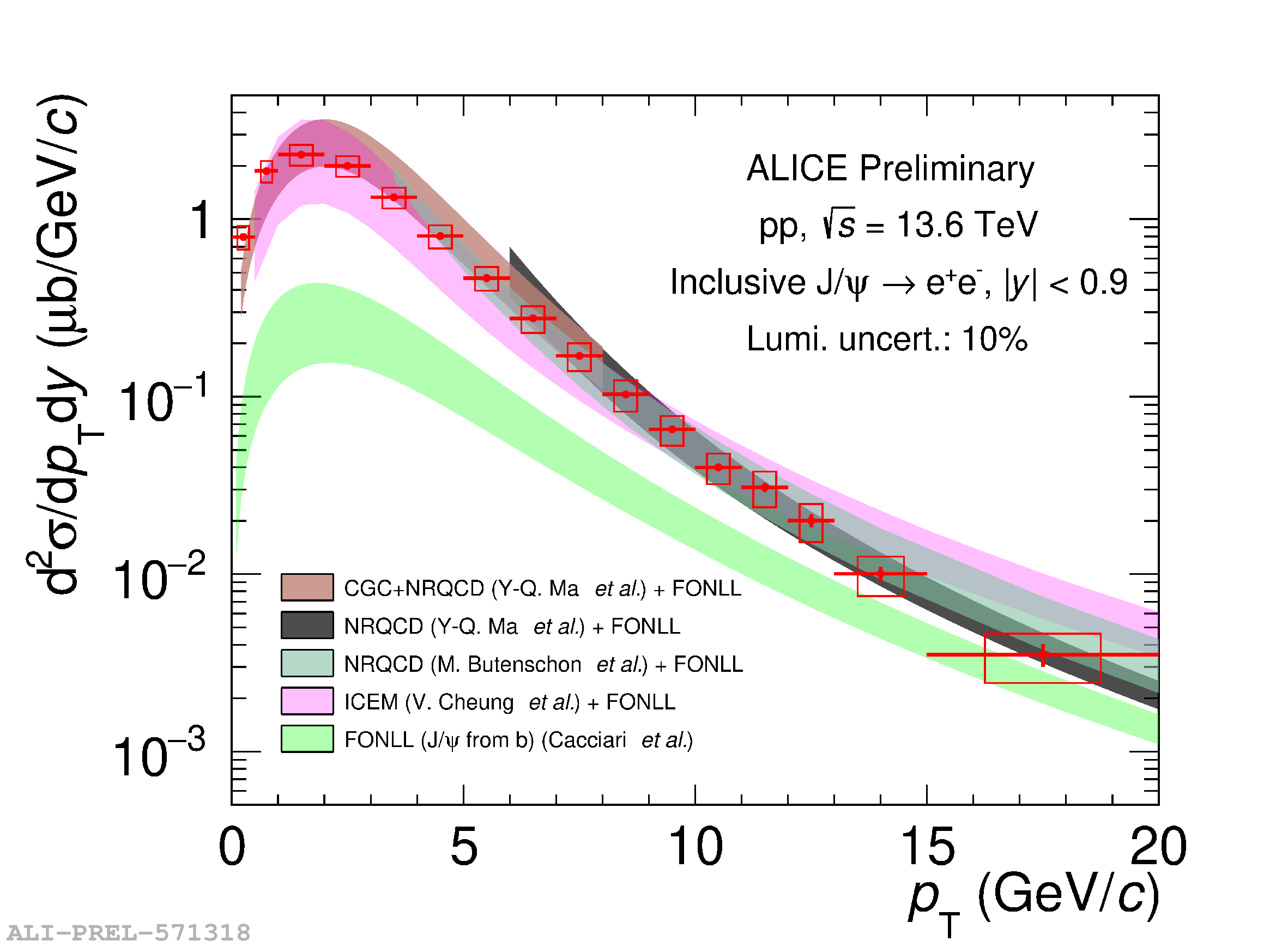}
 	\includegraphics[width=6.3cm]{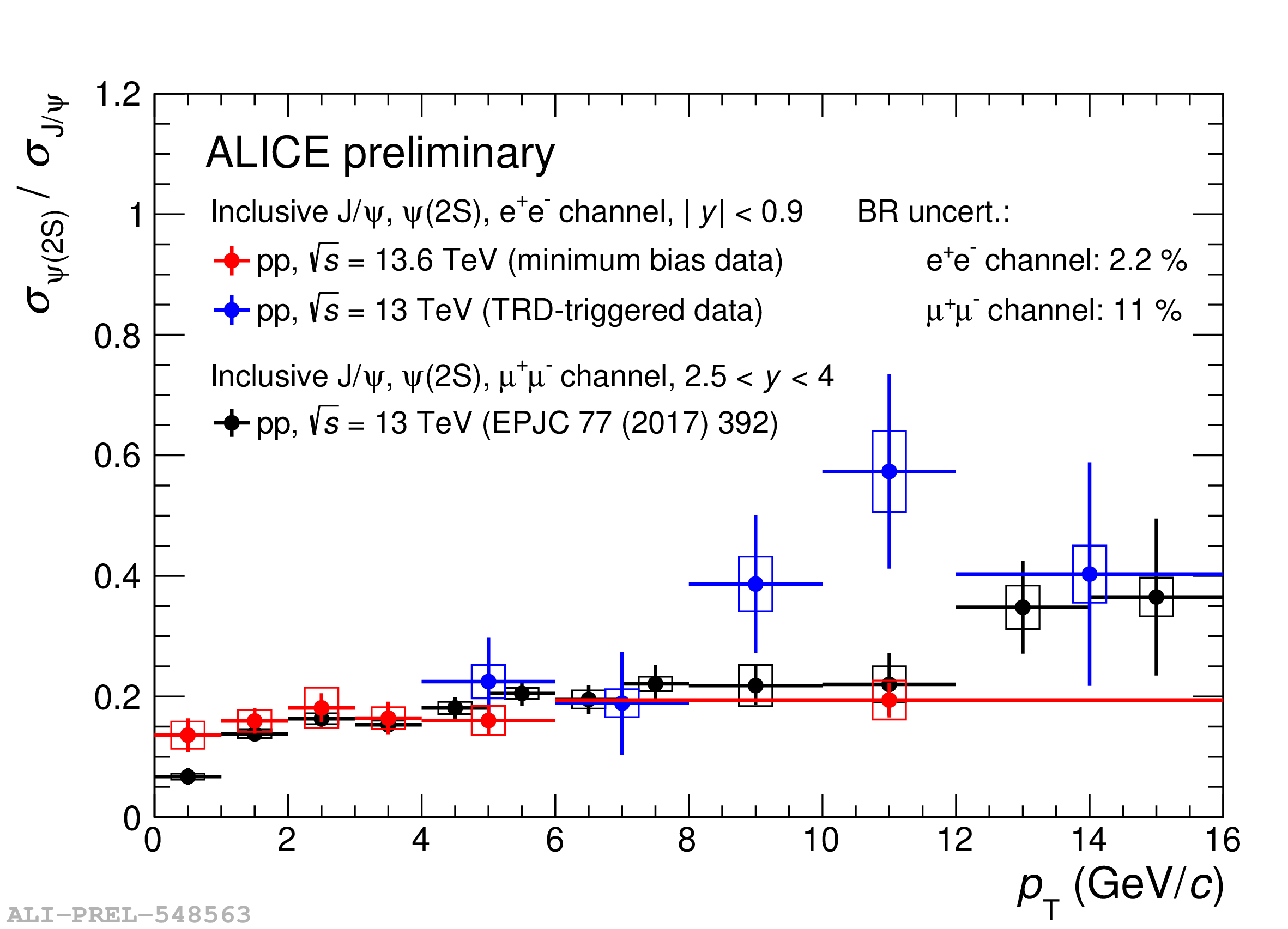}
 	\caption{ (Left) \jpsi production cross section in \pp collisions at \thirteensix. The new results are compared with similar measurements from Run 2 as well as different models calculations, with FONLL accounting for non-prompt contributions. (Right) The \psitwos and \jpsi ratio measured as a function of the \pt. The results are compared with similar measurements from Run 2 at \thirteen, shown at midrapidity (blue) and forward rapidity (black).}
 	\label{DQpp}
 \end{figure}

The left panel of Figure~\ref{DQpp} presents the latest measurement of the \jpsi production cross section in \pp collisions at \thirteensix, which is consistent with results from Run~2. The data are well described by the Improved Color Evaporation Model (ICEM)~\cite{Cheung:2018} and Non-Relativistic QCD (NRQCD) plus the CGC~\cite{Ma:2011,Butenschon:2011,Ma:2014} frameworks, while Fixed-Order Next-to-Leading Logarithm (FONLL) calculations~\cite{Cacciari:1998it} successfully account for the non-prompt \jpsi contribution. The results are compared with theoretical predictions, showing that both CGC+NRQCD~\cite{Zhao:2011cv} and ICEM~\cite{Cheung:2018} approaches provide a good description of the data at low \pt. The right panel of Figure~\ref{DQpp} shows the \psitwos to \jpsi measurement in the \ee decay channel for the first time down to zero \pt at midrapidity, enabled by the significantly increased statistics. The Run3 results compared with Run2 measurements at forward as well as the TRD-triggered data at midrapidity .

\begin{figure}[!hbp]
	\centering
	\includegraphics[width=6.3cm]{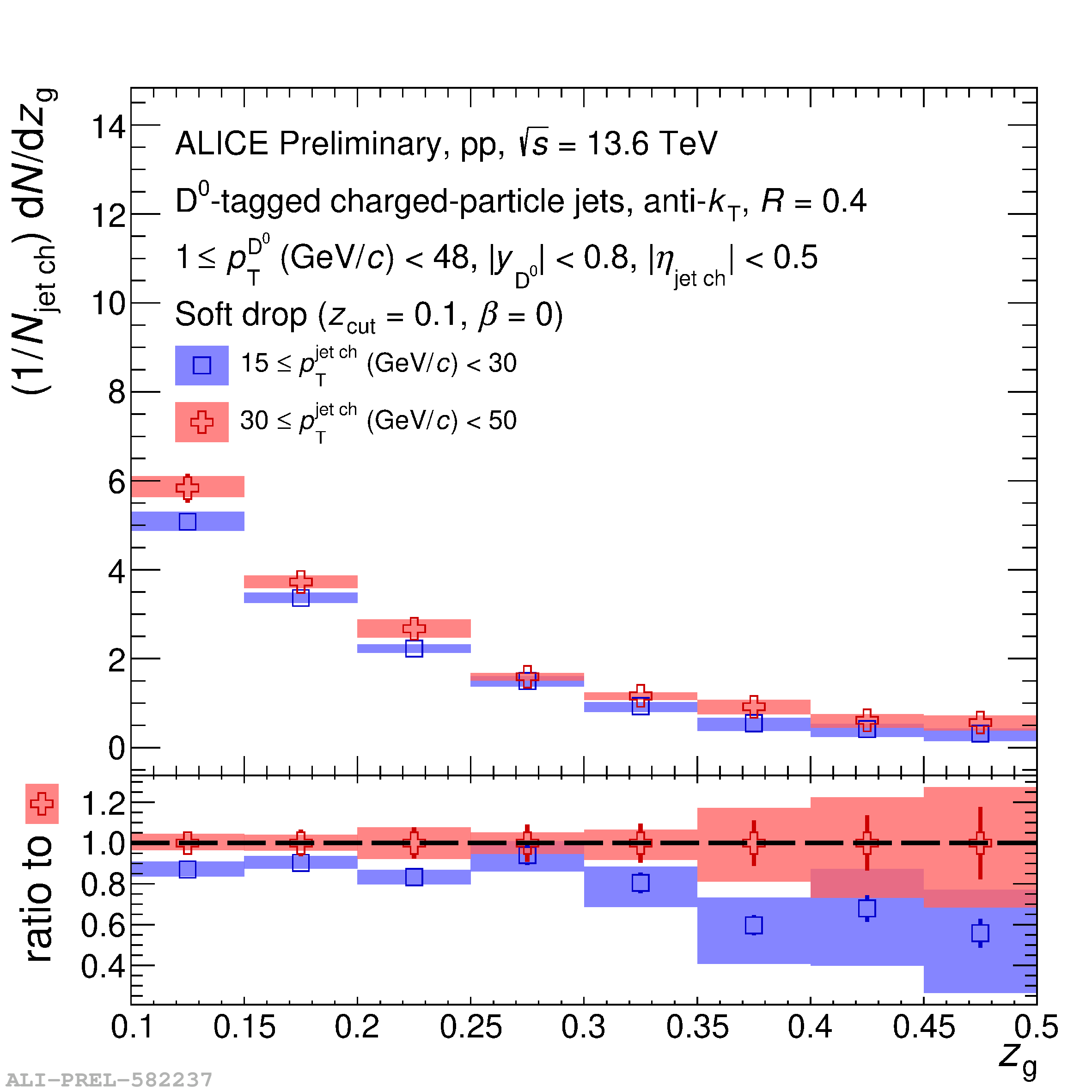}
	\includegraphics[width=6.3cm]{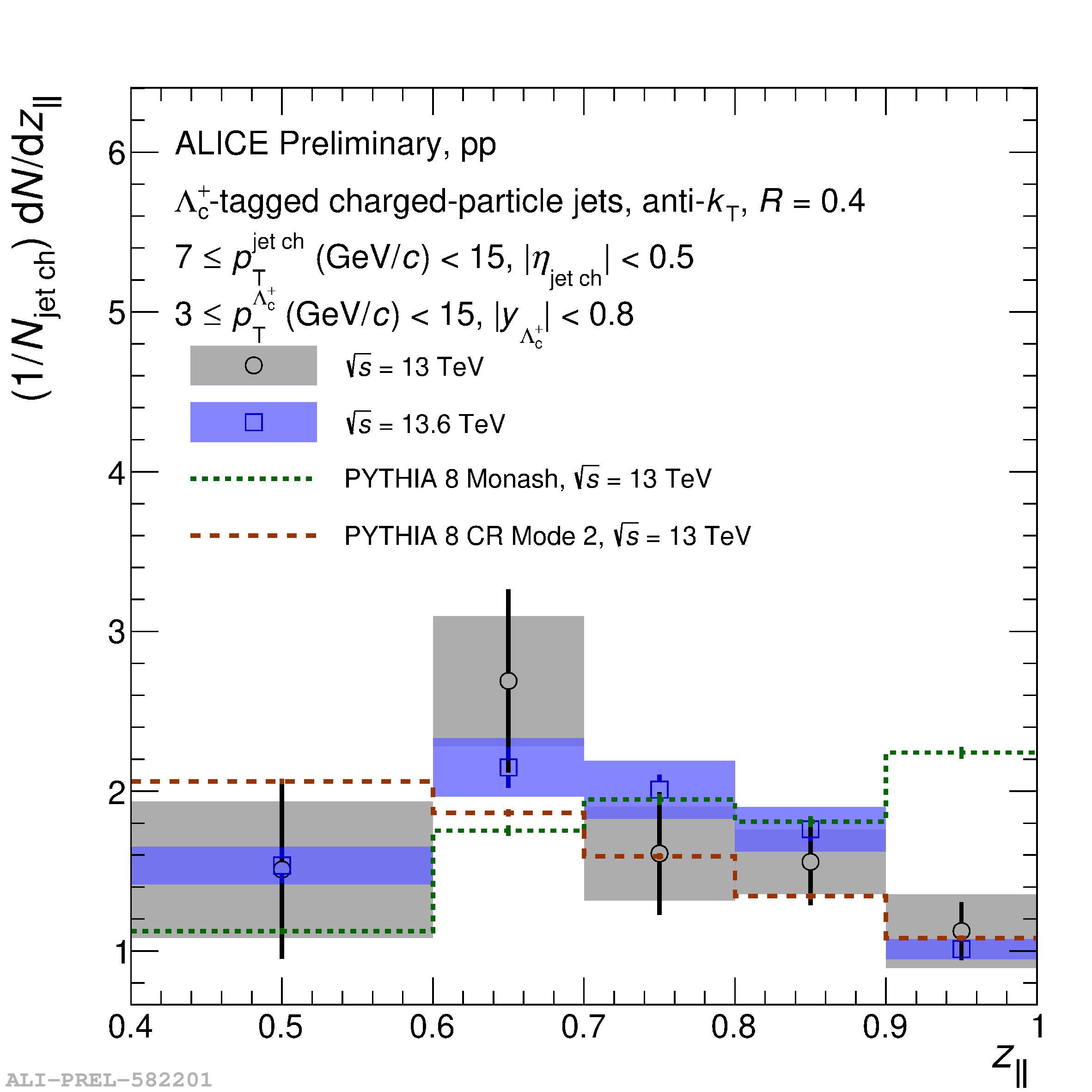}
	\caption{(Left) The $z_g$ distribution of $D^0$-tagged jets for jet transverse momentum ranges of $15 \leq p_{\mathrm{T,jet}}^{\mathrm{ch}} < 30$ GeV/$c$ and $30 \leq p_{\mathrm{T,jet}}^{\mathrm{ch}} < 50$ GeV/$c$ in \pp collisions at $\sqrt{s} = 13.6$ TeV, normalized to the total number of jets. (Right) The $z_{||}$ distribution of $\Lambda_c$-tagged jets in \pp collisions at $\sqrt{s} = 13.6$ TeV, compared with similar measurements in \pp collisions at $\sqrt{s} = 13$ TeV, as well as the PYTHIA 8 calculation.}
	\label{jet-pp}
\end{figure}

Figure~\ref{jet-pp} shows charm-tagged jet substructure in \pp collisions at \thirteensix, exploiting the sizable data sample collected in Run 3. The left panel shows the  distribution of the groomed momentum fraction of the first splitting selected by the Soft Drop grooming algorithm for \dzero-tagged jets~\cite{ALICE:2022mur}, measured in two jet transverse momentum ($p_{\mathrm{T,jet}}$) intervals: 15--30~GeV/$c$ and 30--50~GeV/$c$. This distribution is expected to converge onto the $c \to cg$ splitting function in QCD, allowing for the validation of flavour effects in calculations. Moreover, while $z_g$ remains independent of jet \pt in inclusive jets, a distinct mass-dependent effect is observed in heavy-flavor jets, highlighting the evolution of mass effects with energy in the $c \to cg$ splitting function.

The right panel presents the fraction of the jet longitudinal momentum carried by the \Lc hadron ($z_{||}$) distribution for \Lc-tagged jets, offering direct insights into the fragmentation of charm baryons and highlighting the non-universality of charm baryon hadronization. The significantly improved precision of Run~3 data enables a more differential comparison with earlier \pp measurements at \thirteen and PYTHIA~8 simulations, shedding light on deviations from the relevance of mechanisms beyond the leading-colour approximation and refining our understanding of charm fragmentation dynamics.

\FloatBarrier

\section{Hard probes in the \PbPb collisions}

\begin{figure}[!hbp]
	\centering
	\includegraphics[width=5.5cm]{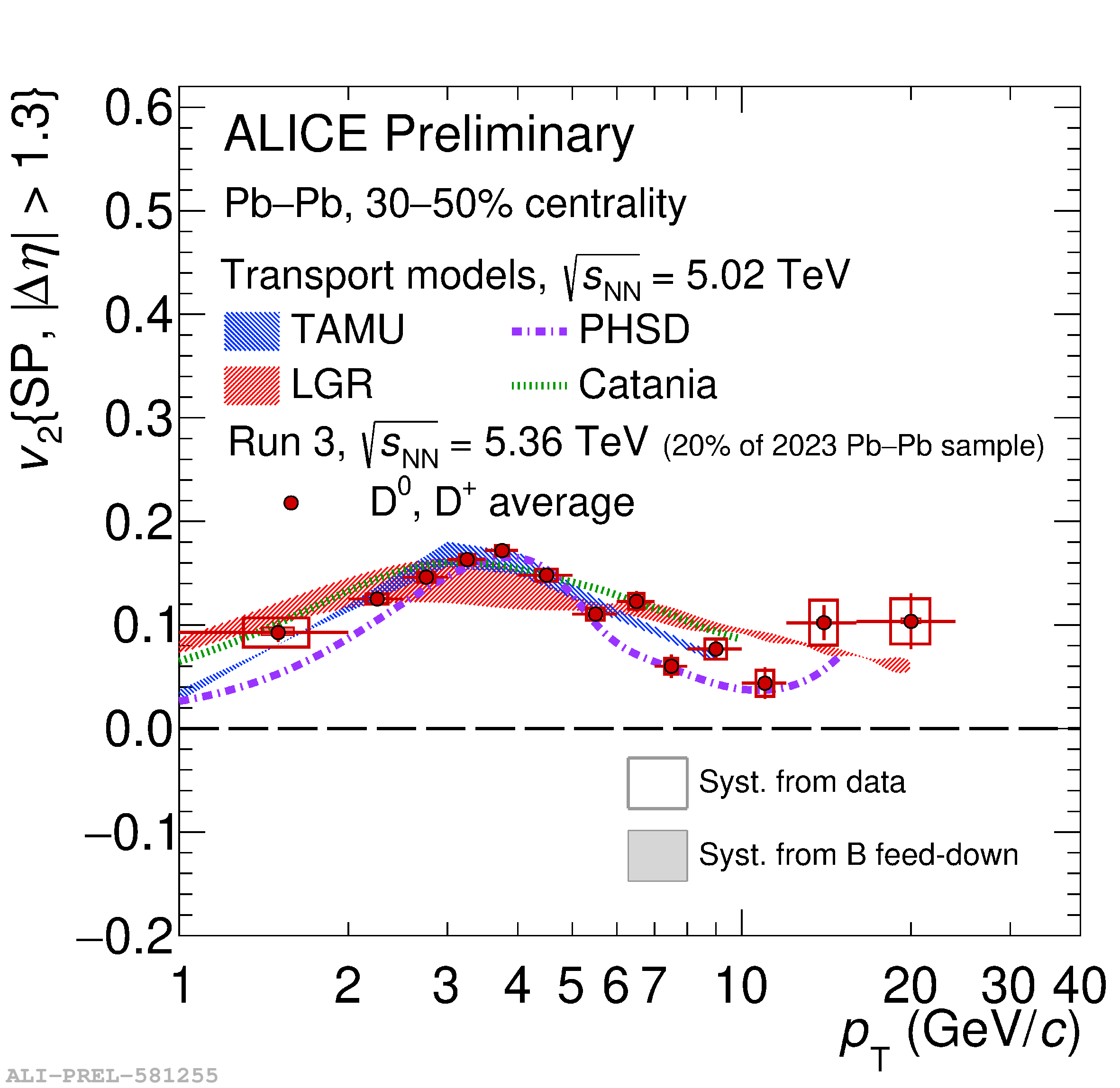}
	\includegraphics[width=7.2cm]{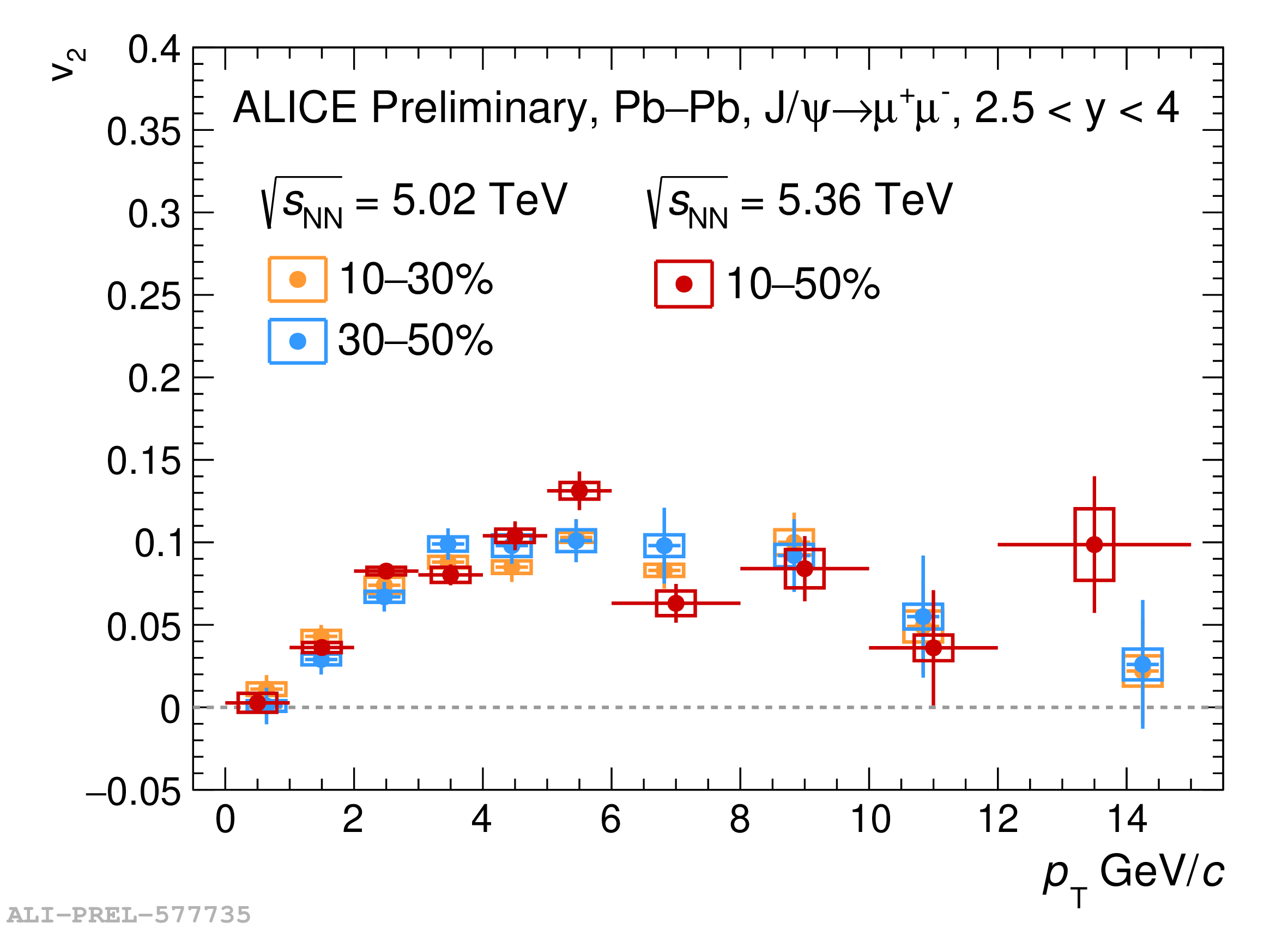}
	  \caption{(Left) The elliptic flow ($v_2$) of averaged prompt \dzero and \dplus mesons in \PbPb collisions at \fivethirtysix. (Right) The inclusive \jpsi meson $v_2$ as a function of \pt at forward rapidity in 10-50\% more central \PbPb collisions at 5.36 TeV compared to the Run 2 published results in 10--30\% and 30--50\%.}	
	\label{v2-pbpb}
\end{figure}

 The collective behavior of the QGP is studied by the elliptic flow ($v_2$). Measuring $v_2$ of heavy flavour hadrons provides important insight into the interaction strength between heavy quarks and the QGP, as well as their transport properties. Figure~\ref{v2-pbpb} shows the $v_2$ of heavy-flavor hadrons in \PbPb collisions at \fivethirtysix, exploiting the large data sample collected in Run~3. The left panel presents the $v_2$ of averaged prompt \dzero and \dplus mesons. A significant positive $v_2$ is observed, which is well described by model calculations from TAMU~\cite{He:2014cla}, PHSD~\cite{Song:2015sfa}, LGR~\cite{Li:2019lex}, and Catania~\cite{Plumari:2017ntm}. These models incorporate in-medium charm quark transport and hadronization mechanisms.

The right panel displays the elliptic flow of inclusive \jpsi as a function of \pt at forward rapidity. The new measurement is consistent with results from Run~2~\cite{Acharya:2020jpsi}, but with improved statistical precision at low \pt, allowing for a more differential study of charmonium flow. A significant non-zero \jpsi $v_2$  is observed, providing evidence for charm quark thermalization in the QGP and reinforcing the role of recombination as a dominant production mechanism at low \pt.

 The nuclear modification factor (\RAA) is important for quantifying the modification of production yields in heavy-ion collisions compared to pp collisions. In the specific case of quarkonia, it is used to describe the color screening, dynamic dissociation, and regeneration effects in QGP.
 
\begin{figure}[!hbp]
	\centering
	\includegraphics[width=7.3cm]{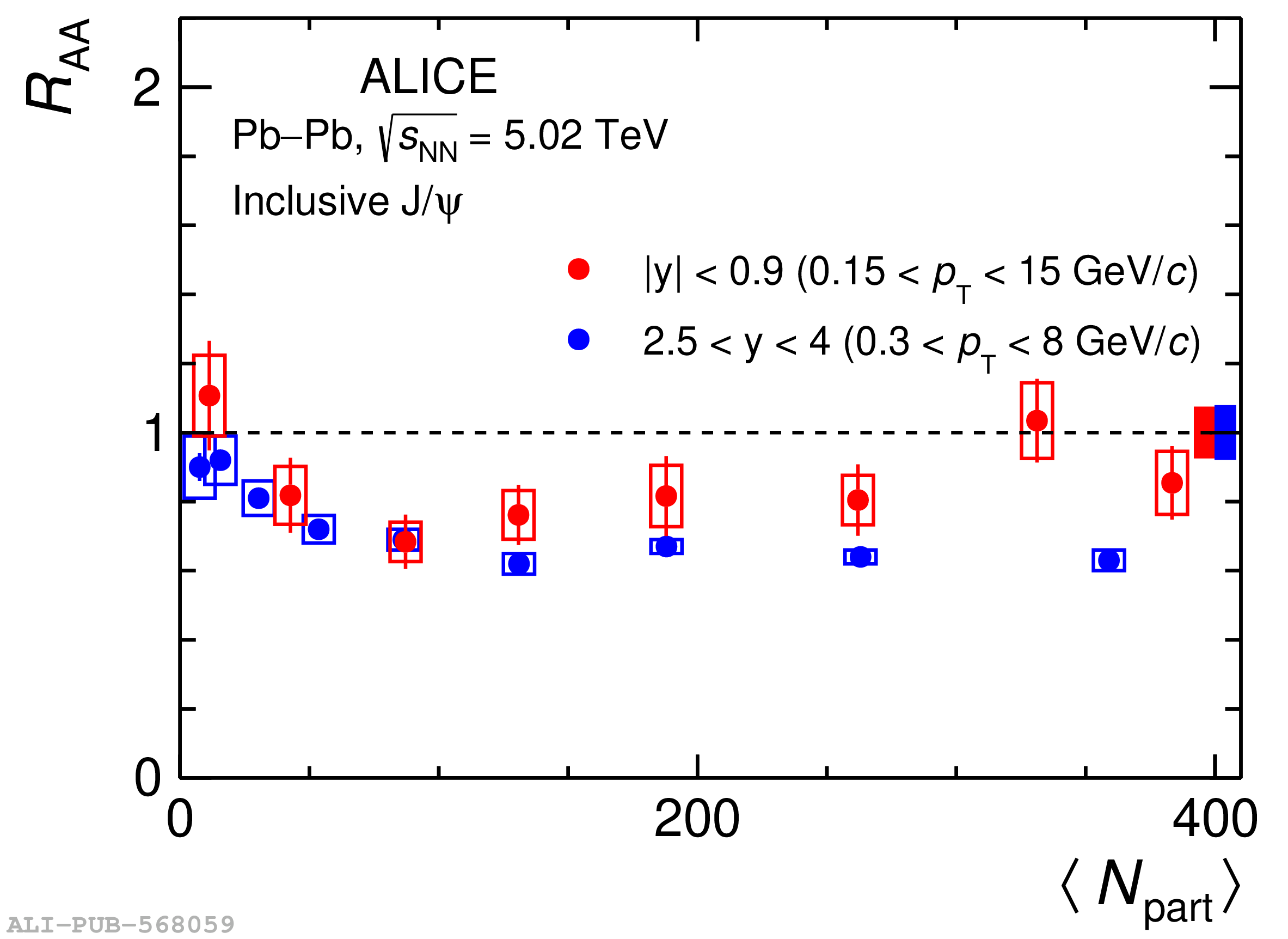}
	\includegraphics[width=5cm]{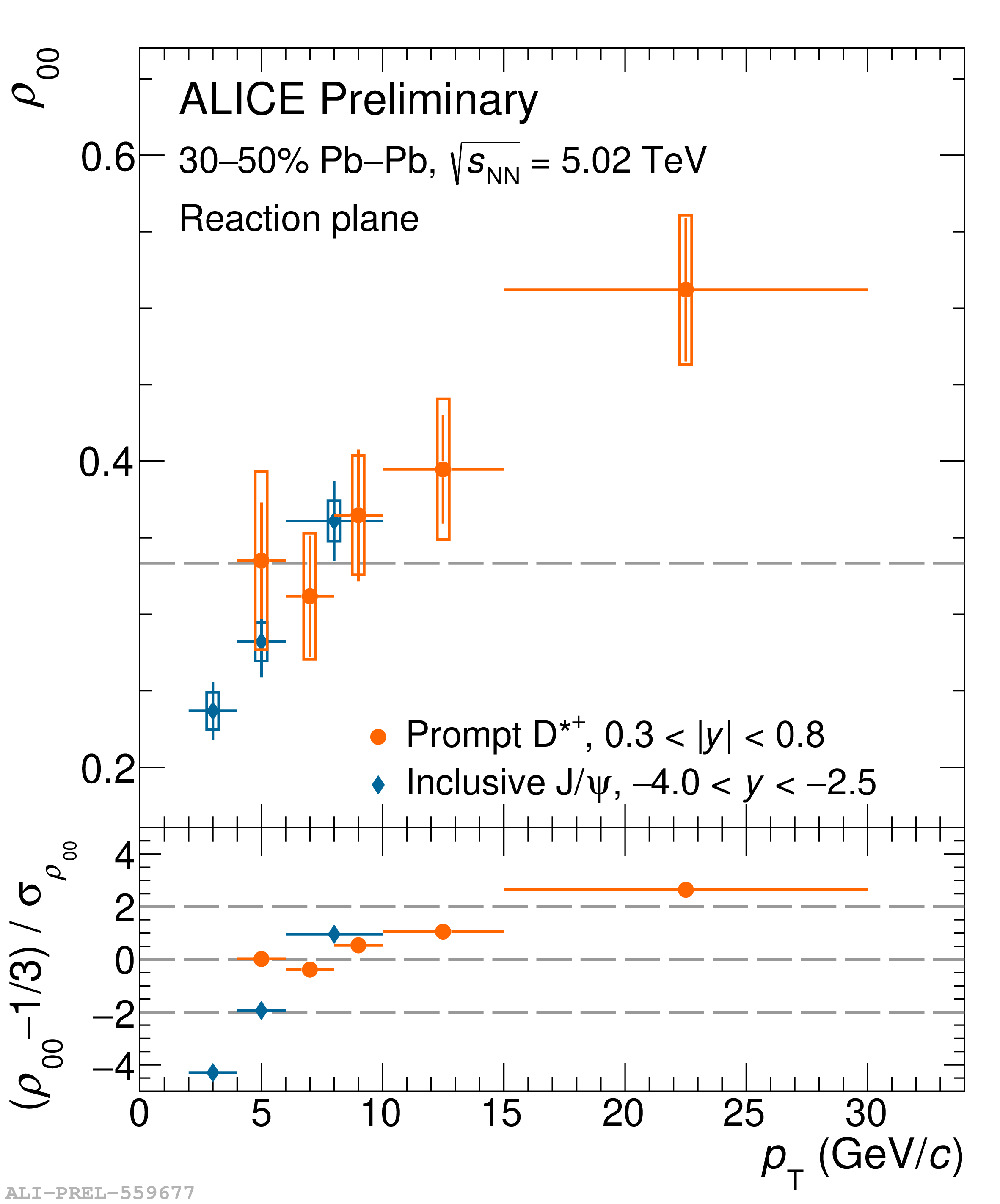}
	  \caption{(Left) The nuclear modification factor \RAA of inclusive \jpsi at midrapidity and forward rapidity as a function of the number of participants in \PbPb collisions at \fivenn. (Right) The polarization parameter $\rho_{00}$ as a function of \pt for inclusive \jpsi and prompt \dstar mesons in \PbPb collisions at \fivenn at 30--50\% centrality interval.}
	\label{RAA-pbpb}
\end{figure}

Figure~\ref{RAA-pbpb} (left) shows the nuclear modification factor \RAA of inclusive \jpsi at midrapidity and forward rapidity as a function of the number of participants in \PbPb collisions at \fivenn~\cite{ALICE:2023gco}. The measurement provides evidence for \jpsi (re)generation in central collisions, with a larger contribution at low \pt and at midrapidity~\cite{Zhou:2014kka}. This observation supports the scenario where recombination of charm quarks in the quark-gluon plasma (QGP) plays a significant role in charmonium production at low momentum. The dependence of \RAA on rapidity and centrality allows for a detailed study of the interplay between suppression and regeneration mechanisms in heavy-ion collisions. 

Extremely strong magnetic fields and large orbital angular momentum are generated in non-central heavy-ion collisions. These induce polarization of charm quarks during the early stages of the collision or  QGP evolutions. The spin alignment of vector mesons, such as the \jpsi, is used to probe these polarization effects. It is quantified by the parameter $\rho_{00}$, which represents the probability of observing the meson in the longitudinal spin state relative to a chosen quantization axis.

The right panel of Figure~\ref{RAA-pbpb} displays the \pt-differential polarization $\rho_{00}$ with respect to the event plane of inclusive \jpsi and prompt \dstar mesons measured in 30-50\% central interval in Pb-Pb collisions at 5.02 TeV. In the absence of polarization, $\rho_{00} = 1/3$ is expected. Deviations from this value signal spin alignment effects due to hadronization or medium interactions. A significant deviation of $\rho_{00}$ from 1/3 is observed at low \pt for \jpsi~\cite{ALICE:2023gco}. The transition toward $\rho_{00}>1/3$ for \dstar at higher \pt is consistent with expectations from two hadronization mechanisms: recombination dominates at low \pt, while fragmentation prevails at high \pt~\cite{Liang:2004ph}. These results offer new insight into heavy-quark spin alignment in the QGP and their hadronization dynamics in \PbPb collisions.

   \FloatBarrier
\section{ALICE future upgrades}

\begin{figure}[H]
	\centering
	\includegraphics[width=6.3cm]{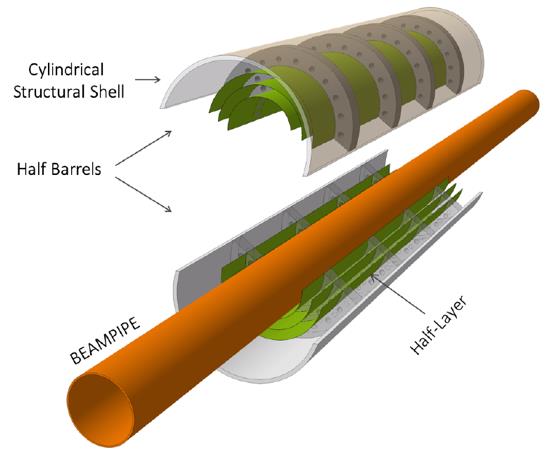}
	\includegraphics[width=6.3cm]{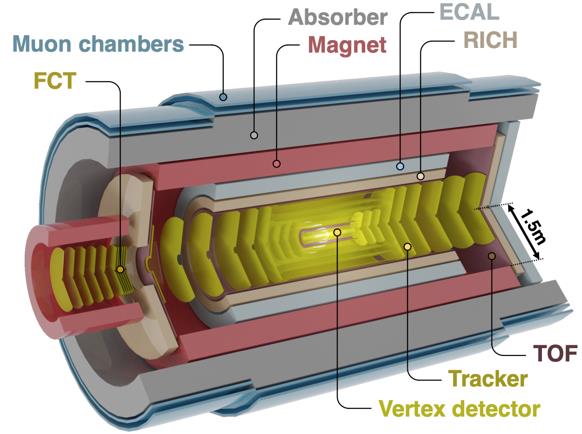}
	\caption{(Left) ALICE ITS3 upgrade for LHC Run~4, designed to enhance tracking performance with a fully silicon-based structure. (Right) ALICE~3 upgrades for LHC Long Shutdown~3 as a next-generation experiment to replace ALICE in LHC Runs~5.}
	\label{alice-upgrade}
\end{figure}

The ALICE experiment is undergoing significant upgrades to enhance its capabilities for Run~4 and beyond. The ITS3 upgrade (left panel of Figure \ref{alice-upgrade}), planned for LHC Run~4, aims to replace the inner layers of ITS2 with a fully silicon-based structure, reducing the material budget and improving tracking precision. The new Monolithic Active Pixel Sensors (MAPS) will be crucial for precision heavy-flavor measurements. Additionally, right panel of Figure \ref{alice-upgrade} shows the FoCal (Forward Calorimeter) upgrade, which will extend ALICE’s reach to the unexplored small Bjorken-$x$ region, enabling studies of direct photons, jets, and heavy quarkonia at forward rapidity.

Beyond Run~4, the ALICE3 project, planned for LHC Runs~5, represents a major overhaul of the experiment~\cite{Adamova:2019vkf}. ALICE3 will feature an all-silicon tracking system with a high-resolution vertex detector and fast readout capabilities, optimizing studies of heavy-flavor production, exotic hadrons, and ultra-soft photon emission. The experiment will also enable high-precision beauty measurements, study multi-charm baryons, and provide insights into thermal radiation and QGP dynamics. These upgrades will ensure that ALICE remains at the forefront of heavy-ion physics, delivering unprecedented precision in exploring the QGP and hadronization mechanisms in high-energy nuclear collisions.

\section{Summary}

These proceedings reported the selected ALICE highlight results on hard probes, primarily based on Run~3 data. It also discusses the performance of the upgraded ALICE Run~3 detectors, future plans for Run~4, and the long-term upgrade projects for ALICE 3 during LS3.

\section*{Acknowledgement}
The author is supported in part by the Ministry of Science and Technology of China under Grant No.~2024YFA1610803, the NSFC under grant No.~12475146 and ~2061141008, the Fundamental Research Funds for the Central Universities.

 \FloatBarrier

\bibliography{reference.bib}

\end{document}